\documentclass[final,5p,twocolumn]{elsarticle}  

\usepackage{graphicx} 
\usepackage{tikz}
\usepackage{pgfplots,pgfplotstable}
\pgfplotsset{compat=1.18}
\usepackage{subcaption} 
\usepackage{multirow}
\usepackage{booktabs}
\usepackage{amsmath}
\usepackage{textcomp}
\usepackage{comment}
\usepackage{xcolor}
\usepackage{fancybox}
\usepackage[normalem]{ulem}
\usepackage{makecell}
\usetikzlibrary{shapes,arrows}
\usepackage{empheq}
\usepackage{stfloats}
\usepackage[hyphens]{url}
\usepackage[hidelinks]{hyperref}
\hypersetup{colorlinks=true,linkcolor=black,citecolor=black,filecolor=black,urlcolor=black,breaklinks=true}
\usepackage{caption}
\usepackage{color}
\usepackage{cancel}
\usepackage{array}
\usepackage[export]{adjustbox}
\usepackage[normalem]{ulem}
\usepackage{amsfonts}
\usepackage{float}
\usepackage{amsmath,amssymb,amsfonts}
\usepackage{algorithmic}
\usepackage{svg}
\usepackage{xcolor}
\usepackage{multicol}
\usepackage{enumitem}
\usepackage{flushend}
\usepackage[dvipsnames]{xcolor}
\usepackage{balance}

\begin{document}

\begin{frontmatter}

\title{On the Impact of Voltage Unbalance on\\Distribution Locational Marginal Prices}

\author[inst1]{Alireza~Zabihi\corref{cor1}}
\ead{alireza.zabihi@upm.es}

\author[inst1]{Luis~Badesa}
\ead{luis.badesa@upm.es}

\author[inst1]{Araceli~Hernández}
\ead{araceli.hernandez@upm.es}

\cortext[cor1]{Corresponding author}

\affiliation[inst1]{organization={Universidad Politécnica de Madrid},
            city={Madrid},
            country={Spain}}

\begin{abstract}
Voltage unbalance (VU) is becoming a critical operational challenge in modern distribution networks due to the increasing penetration of single-phase loads and distributed energy resources. VU affects power quality, accelerates asset degradation, and increases system losses, making its efficient management essential.
Traditional grid codes address unbalance via disparate hard limits on various indices thresholds that differ across standards, offer no dynamic economic incentive and undermine optimality. 
This paper introduces a `soft limit' framework that embeds VU penalties into a three-phase optimal power flow formulation to reflect the economic impact of asset lifetime reduction under unbalanced voltages. The approach yields unbalance-aware Distribution Locational Marginal Prices (DLMPs), providing dynamic price signals tied directly to power quality. A new analytical decomposition of DLMPs is developed to isolate energy, loss, congestion, and unbalance components, enabling transparent interpretation of their drivers.
Case studies conducted on two benchmark networks demonstrate the effectiveness and practical value of the proposed method. The results indicate that unbalance penalties reshape nodal prices, produce unexpected phase-level effects, and even allow scenarios where added load reduces unbalance and lowers costs, while providing planners and market designers with actionable insights to balance investment, operation, and power quality in modern distribution systems.
\end{abstract}

\begin{keyword}
Distribution locational marginal prices \sep Power quality \sep Three-phase optimal power flow \sep Voltage unbalance
\end{keyword}

\end{frontmatter}


\section*{Nomenclature}

\subsection*{Indices and Sets}
\begin{description}[leftmargin=1.5cm, labelwidth=1.5cm, labelsep=0.0cm, align=left]
    \item[$\{i, j\},\,L$] Index, Set of power links.
    \item[$i,\,N$] Index, Set of three-phase nodes.
    \item[$P_{i}^c$] Set of three-phase power consumed at node $i$.
    \item[$v_i$] Set of three-phase voltages of node $i$.
    \item[$\varphi,\,\Phi$] Index, Set of electrical phases.
\end{description}

\subsection*{Constants and Functions}
\begin{description}[leftmargin=0.9cm, labelwidth=0.9cm, labelsep=0.0cm, align=left]
    \item[$C_{i_{\varphi}}^{g/c}$] Cost function of generators/consumers connected to node $i$ on phase $\varphi$ (€/h).
    \item[$C_i^{VU}$] Cost function of voltage unbalance penalty on node $i$ (€/h).
    \item[$\textrm{c}^{\textrm{VU}}_{\textrm{i}}$] Marginal penalty of voltage unbalance on node $i$ (€/h.\%).
\end{description}
\begin{description}[leftmargin=1.8cm, labelwidth=1.8cm, labelsep=0.0cm, align=left]
    \item[$f(v_i)$] Function that defines the voltage unbalance index in the grid (\%).
    \item[$\textrm{P}_{\textrm{i}_{\varphi}}^{\textrm{g-}}/\textrm{P}_{\textrm{i}_{\varphi}}^{\textrm{g+}}$] Minimum/maximum active power limits of generators (kW).
    \item[$\textrm{Q}_{\textrm{i}_{\varphi}}^{\textrm{g-}}/\textrm{Q}_{\textrm{i}_{\varphi}}^{\textrm{g+}}$] Minimum/maximum reactive power limits of generators (kvar).
    \item[$\bar{\textrm{s}}_{\textrm{i}\textrm{j}_{\varphi}}$] Thermal limit of the links in each phase (kVA).
    \item[$\textrm{V}^{-}/\textrm{V}^{+}$] Minimum/maximum voltage magnitude limits (V).
    \item[$\overline{\textrm{V.U}.}$] Maximum admissible level of voltage unbalance (\%).
\end{description}

\subsection*{Primal Variables}
\begin{description}[leftmargin=0.9cm, labelwidth=0.9cm, labelsep=0.0cm, align=left]
    \item[$I_{ij,\varphi}$] Current flowing between node $i$ and $j$ in phase $\varphi$ (A).
    \item[$P_{i_{\varphi}}^{g/c}$] Total active power produced/consumed at node $i$ in phase $\varphi$ (kW).
    \item[$p_{ij_{\varphi}}$] Active power transferred from node $i$ to node $j$ through the $\{i, j\}$ link in phase $\varphi$ (kW).
    \item[$Q_{i_{\varphi}}^{g/c}$] Total reactive power produced/consumed by generators at node $i$ in phase $\varphi$ (kvar).
    \item[$q_{ij_{\varphi}}$] Reactive power transferred from node $i$ to node $j$ through the $\{i, j\}$ link in phase $\varphi$ (kvar).
    \item[$v_{i, \varphi}$] Voltage of node $i$ in phase $\varphi$ (V).
\end{description}

\subsection*{Dual Variables}
\begin{description}[leftmargin=1.6cm, labelwidth=1.6cm, labelsep=0.0cm, align=left]
    \item[$\delta_{i, \varphi}^{g-}/\delta_{i, \varphi}^{g+}$] Multipliers of the minimum/maximum active power output of the generators.
    \item[$\eta_{i j, \varphi}$] Multiplier of the congestion limit.
    \item[$\phi_{i, \varphi}^p / \phi_{i, \varphi}^q$] Multipliers of the nodal real/reactive power balance.
    \item[$\psi_i$] Multiplier of the maximum admissible level of voltage unbalance.
    \item[$\sigma_{i, \varphi}^{-}/\sigma_{i, \varphi}^{+}$] Multipliers of voltage magnitude limits.
    \item[$\theta_{i, \varphi}^{g-}, \theta_{i, \varphi}^{g+}$] Multipliers of the minimum/maximum reactive power output of the generators.
\end{description}

\section{Introduction}
Electric power distribution is undergoing a profound transformation driven by the large-scale integration of single-phase technologies such as electric vehicles (EVs) and distributed energy resources (DERs). While these developments support decarbonization and enhance system resilience, they introduce persistent phase-to-phase imbalances that lead to voltage unbalance across the network. This condition degrades power quality, increases losses, and accelerates wear on three-phase equipment and industrial machinery \cite{woolley2012statistical}, \cite{ma2020review}, imposing significant costs on utilities and end users \cite{PERERA202349}. As the penetration of single-phase devices continues to rise, voltage unbalance (VU) is expected to intensify, posing a serious threat to the efficient operation of distribution networks and the equitable allocation of costs among users and phases. Understanding its impact on network losses \cite{LoadUnbalanceLosses2019}, capital expenditures, and locational electricity prices is essential; without such insights, market mechanisms risk misallocating costs, distorting investment signals, and undermining economic efficiency \cite{liu2022using}.

Distribution Locational Marginal Prices (DLMP) provide a principled framework for allocating operational costs to specific nodes and phases \cite{liu2017distribution}. Derived from the dual variables of an AC Optimal Power Flow (AC-OPF) formulation, DLMPs can be decomposed into interpretable components—such as energy, losses, and congestion—that offer actionable signals for network planning, DER siting, and market settlement \cite{heydt2012pricing}, \cite{papavasiliou2017analysis}. 
Motivated by these properties, a substantial body of work has extended DLMP theory to distribution settings and market operation, emphasizing DLMP as a marginal cost-based mechanism that supports cost causation, efficient DER participation, and decentralized or competitive clearing structures \cite{wang2022dlmp, sabillon, ruan2025distribution}.

In parallel, existing literature has examined unbalanced three-phase DLMPs, showing how phase dependent losses, congestion, voltage constraints, and imbalance related effects influence nodal price formation \cite{papavasiliou2017analysis, ElsJ1}. Probabilistic and scenario-based formulations further demonstrate how uncertainty propagates into phase-level DLMP components under unbalanced operating conditions \cite{Saeed}.

However, despite these advances, the prevailing three-phase DLMP literature typically treats unbalance primarily as a structural feature of the feeder and focuses on accurately pricing energy, losses, congestion, and voltage magnitude constraints across phases \cite{ElsJ1, Saeed, ACOPF_Linearized}. As a result, the power quality cost of VU linked to increased losses, negative sequence effects, and accelerated asset degradation remains largely external to the market objective and is not explicitly reflected as a distinct marginal component in DLMP signals.

At the regulatory level, VU is controlled through hard limits on indices defined by standards such as IEC/TR~61000-3-13 \cite{IEC} and IEEE~1159 \cite{IEEE1159}. Although these limits ensure equipment protection, they do not provide dynamic economic incentives and can be restrictive or non-aligned with cost efficient operation in DER-rich environments. Engineering evidence that even mild unbalance can yield disproportionate negative sequence currents, elevated losses, and accelerated insulation degradation \cite{vonJouanne2001} motivates the need for market signals that internalize the operational cost of VU rather than treating it solely as a pass/fail constraint.
Meanwhile, recent contributions in unbalanced network operation and pricing \cite{ElsJ2, ElsJ3} further reinforce the need for three-phase-aware economic signals that can be used consistently in both operational control and planning.

Incorporating VU into DLMP decomposition presents two primary challenges. First, accurately modeling unbalance requires a full three-phase AC-OPF, which is inherently nonconvex and computationally demanding \cite{taylor2015convex, ACOPF_DLMP}. To manage complexity, many studies rely on linearized \cite{ACOPF_Linearized} or approximate models \cite{Saeed, ACOPF_moment}, which may not capture the nuanced impacts of unbalance. Second, current grid codes typically address unbalance through hard constraints on various unbalance indices \cite{Zabihi2025Voltage}. These indices—and their permissible thresholds—vary widely across jurisdictions \cite{IEC, IEEE1159}, leading to inconsistent reliability margins and regulatory fragmentation \cite{girigoudar2020impact}, \cite{churkin2024quantifying}. While \textit{hard limits} ensure operational safety, they do not support optimization of social welfare or cost efficiency \cite{Badesa:2023}, \cite{Ding2014_ExactPenalty}.

In the broader OPF and market design literature, it is important to note that \textit{soft limits} and penalty-based OPF methods are not new in general \cite{Ding2014_ExactPenalty}. The open gap addressed here is specifically: how to (i) embed voltage unbalance as a monetized power quality externality within a three-phase social welfare problem, and (ii) obtain a DLMP decomposition whose dual variables yield an explicit and interpretable unbalance price component that can be used for market settlement and planning.

This paper addresses these gaps by embedding voltage unbalance directly into a social welfare maximization framework. Instead of imposing strict constraints, penalty terms are applied to node-level unbalance factors and incorporated into the generation cost function. This integration accounts for the impact of voltage unbalance, which is recognized as a power quality violation that can reduce the operational lifetime of assets exposed to such conditions.
Although small VUF values may be considered acceptable by existing standards, this does not imply that their impact is negligible. Even low levels of VU can affect the lifetime of grid assets, and the \textit{soft limits} formulation explicitly accounts for these effects. There are several studies attempting to find impact of VU on lifetime of assets~\cite{Jalilian-Induction, Molzahn_rep}.
This treatment aligns with exact penalty approaches for OPF \cite{Ding2014_ExactPenalty} and enhances the interpretability and economic relevance of DLMPs in increasingly unbalanced distribution systems.
This approach yields a unified optimization problem whose dual variables naturally reflect the penalization cost of unbalance within the DLMP structure \cite{Penalty1, Penalty2}. The resulting decomposition explicitly isolates the cost of unbalance, offering clear economic signals for infrastructure upgrades, EV charging coordination, DER placement, and market pricing strategies.  

In contrast to existing three-phase DLMP formulations that primarily price the feasibility of voltage magnitudes and network constraints, the proposed framework prices power quality degradation itself through a marginal unbalance component, thereby strengthening the link between distribution market outcomes and asset health and equipment cost externalities.

The main contributions of this work can be presented as:

\begin{enumerate} 
\item A novel three-phase social-welfare maximization framework is introduced, which incorporates cost-based penalties for nodal voltage unbalance in place of rigid code-based limits, while maintaining the persistence of appropriate economic signals. 
\item Intuitive expressions for DLMP are derived, enabling a clear separation of energy, loss, congestion, and, for the first time, voltage unbalance components. 
\item Voltage unbalance effects are embedded into DLMPs via exact penalty terms, producing clear marginal cost signals that reflect power quality degradation and support economically efficient operational and planning decisions.
\end{enumerate}

The remainder of this paper is organized as follows. Section~\ref{sec:2-A} presents the mathematical model and the decomposition of DLMPs. Section~\ref{math_math} discusses the voltage unbalance components in DLMPs. Section~\ref{sec:test} provides the case studies and analysis. Finally, Section~\ref{sec:conclusion} concludes the paper.

\section{Formulations and Decomposition}
\label{sec:formulations}
In this section, the formulation and decomposition of the unbalance-aware OPF problem are initially presented for both the \textit{hard limit} and \textit{soft limit} approaches.
To derive the decomposition of DLMP under both formulations and to evaluate the cost of increasing demand, duality theory and the Karush–Kuhn–Tucker (KKT) conditions are employed.
The second subsection focuses specifically on the components related to voltage unbalance, where a detailed decomposition is derived.

\subsection{Optimal Power Flow Formulations and Unbalance-aware DLMP Decompositions}
\label{sec:2-A}
The general formulations of three-phase OPF or maximization of social welfare \cite{papavasiliou2017analysis} using \textit{hard} and \textit{soft} limits are as follows. The dual variables corresponding to each constraint are given on their left side in Greek letters. 
The first formulation uses the conventional method, where the upper limit of voltage unbalance at each node is enforced as a constraint, or \textit{hard limit} \cite{churkin2024quantifying}. In contrast, the second formulation treats voltage unbalance as a penalty term in the objective function, effectively applying it as a \textit{soft limit}.

\subsubsection{Hard limit (Voltage unbalance as a constraint)}
In this case, the grid code thresholds are imposed as constraints in the problem.
\begin{equation}
\label{eq:social_welfare1}
\text { (I): } \max \sum_{\substack{i \in N \\
\varphi \in \Phi}} C_{i_{\varphi}}^c\left(P_{i_{\varphi}}^c\right)-C_{i_{\varphi}}^g\left(P_{i_{\varphi}}^g\right) 
\end{equation}

subject to:
\begin{align}
\phi_{i, \varphi}^p &:\; \sum_{j:\{i, j\} \in L} p_{i j_{\varphi}} = P_{i_{\varphi}}^g - P_{i_{\varphi}}^c
\label{eq:P_balance1} \\
\phi_{i, \varphi}^q &:\; \sum_{j:\{i, j\} \in L} q_{i j_{\varphi}} = Q_{i_{\varphi}}^g - Q_{i_{\varphi}}^c
\label{eq:Q_balance1} \\
\sigma_{i, \varphi}^{-},\, \sigma_{i, \varphi}^{+} &:\; \text{V}^{-} \le \Vert v_{i, \varphi} \Vert \le \text{V}^{+}
\label{eq:Vmag} \\
\delta_{i, \varphi}^{g-},\, \delta_{i, \varphi}^{g+} &:\; \textrm{P}_{\textrm{i}_{\varphi}}^{\textrm{g-}} \le P_{i_{\varphi}}^g \le \textrm{P}_{\textrm{i}_{\varphi}}^{\textrm{g+}}
\label{eq:Pg} \\
\theta_{i, \varphi}^{g-},\, \theta_{i, \varphi}^{g+} &:\; \textrm{Q}_{\textrm{i}_{\varphi}}^{\textrm{g-}} \le Q_{i_{\varphi}}^g \le \textrm{Q}_{\textrm{i}_{\varphi}}^{\textrm{g+}}
\label{eq:Qg} \\
\eta_{i j, \varphi} &:\; p_{i j_{\varphi}}^2 + q_{i j_{\varphi}}^2 \le \bar{\textrm{s}}_{\textrm{i} \textrm{j}_{\varphi}}^2,\ \{i, j\} \in L
\label{eq:Thermal1} \\
\psi_i &:\; f\left(v_i\right) \le \overline{\textrm{V.U}.}
\label{eq:V.U._constraint}
\end{align}

The objective of~(\ref{eq:social_welfare1}) is to maximize social welfare. Constraints~(\ref{eq:P_balance1}) and~(\ref{eq:Q_balance1}) represent the active and reactive nodal power balances. Constraint~(\ref{eq:Vmag}) enforces the lower and upper limits on voltage magnitudes. Constraints~(\ref{eq:Pg}) and~(\ref{eq:Qg}) define the bounds for active and reactive power generation. Constraint~(\ref{eq:Thermal1}) ensures thermal limits of the distribution lines are respected. Finally, constraint~(\ref{eq:V.U._constraint}) sets the upper limit for the voltage unbalance level at each node, \(f\) is the function that defines the voltage unbalance metric in the grid.

To determine the marginal cost of power consumption in a particular node and phase, it is necessary to construct the Lagrangian function and apply the corresponding stationary conditions:

\begin{equation}
\begin{aligned}
\mathcal{L} = & \sum_{\substack{i,j \in N \\ \{i,j\} \in L \\ \varphi \in \Phi}} \Bigg[ C_{i_{\varphi}}^g (P_{i_{\varphi}}^g) - C_{i_{\varphi}}^c (P_{i_{\varphi}}^c) \\
& + {\phi}_{i,{\varphi}}^p \left( P_{i_{\varphi}}^c - P_{i_{\varphi}}^g + \sum_{j: \{i,j\} \in L} p_{ij_{\varphi}} \right) \\
& + {\phi}_{i,{\varphi}}^q \left( Q_{i_{\varphi}}^c - Q_{i_{\varphi}}^g + \sum_{j: \{i,j\} \in L} q_{ij_{\varphi}} \right) \\
& + \sigma_{i,{\varphi}}^+ (\Vert v_{i,{\varphi}} \Vert - \textrm{V}^+) + \sigma_{i,{\varphi}}^- (\textrm{V}^- - \Vert v_{i,{\varphi}} \Vert) \\
& + \delta_{i,{\varphi}}^{g+} (P_{i_{\varphi}}^g - \textrm{P}_{\textrm{i}_{\varphi}}^{\textrm{g+}}) + \delta_{i,{\varphi}}^{g-} (\textrm{P}_{\textrm{i}_{\varphi}}^{\textrm{g-}} - P_{i_{\varphi}}^g) \\
& + \theta_{i,{\varphi}}^{g+} (Q_{i_{\varphi}}^g - \textrm{Q}_{\textrm{i}_{\varphi}}^{\textrm{g+}}) + \theta_{i,{\varphi}}^{g-} (\textrm{Q}_{\textrm{i}_{\varphi}}^{\textrm{g-}} - Q_{i_{\varphi}}^g) \\
& + \eta_{i j, \varphi} (p_{i j, \varphi}^2 + q_{i j, \varphi}^2 - \bar{s}_{i j, \varphi}^2) + \psi_i (f(v_i) - \overline{\textrm{V.U}.}) \Bigg]
\end{aligned}
\label{L_1}
\end{equation}
Where its corresponding stationarity condition with respect to $P_{n_{\varphi}}^c$ implies that $\frac{\partial \mathcal{L}}{\partial P_{n_{\varphi}}^c} = 0$, so the DLMP of active power for node $n$ and on phase $\varphi$ is equal to:
\begin{equation}
\label{Cons}
\begin{aligned}
\text{DLMP}_{n,{\varphi}} & = {\phi}_{n,{\varphi}}^p \sum_{\{n,j\} \in L} \left( 1 + \frac{\partial p_{nj_{\varphi}}}{\partial P_{n_{\varphi}}^c} \right) + \sum_{\substack{i \neq n \\ \{i,j\} \in L}} {\phi}_{i,{\varphi}}^p \frac{\partial p_{ij_{\varphi}}}{\partial P_{n_{\varphi}}^c} \\
& + {\phi}_{n,{\varphi}}^q \sum_{\{n,j\} \in L} \frac{\partial q_{nj_{\varphi}}}{\partial P_{n_{\varphi}}^c} + \sum_{\substack{i \neq n \\ \{i,j\} \in L}} {\phi}_{i,{\varphi}}^q \frac{\partial q_{ij_{\varphi}}}{\partial P_{n_{\varphi}}^c} \\
& + \sigma_{n,{\varphi}}^+ \frac{\partial \Vert v_{n,{\varphi}} \Vert}{\partial P_{n_{\varphi}}^c} - \sigma_{n,{\varphi}}^- \frac{\partial \Vert v_{n,{\varphi}} \Vert}{\partial P_{n_{\varphi}}^c} \\
& + 2 \sum_{\{i,j\} \in L} \eta_{ij,{\varphi}} \left( p_{ij_{\varphi}} \frac{\partial p_{ij_{\varphi}}}{\partial P_{n_{\varphi}}^c} + q_{ij_{\varphi}} \frac{\partial q_{ij_{\varphi}}}{\partial P_{n_{\varphi}}^c} \right) \\
& + \psi_n \frac{\partial f(v_n)}{\partial P_{n_{\varphi}}^c}
\end{aligned}
\end{equation}

The term reflecting the impact of the voltage unbalance constraint is the last term of equation (\ref{Cons}), where $\frac{\partial f(v_n)}{\partial P_{n_{\varphi}}^c}$ represents the voltage unbalance changes in node $n$ due to small changes in demand in the same node and phase $\varphi$. This function is explored in detail in subsection \ref{math_math}.

\subsubsection{Soft limit (Voltage unbalance as a penalty term)}
Here, the \textit{hard limit} is relaxed and the cost of voltage unbalance is considered as a penalty term in the objective function.
\begin{equation}
\label{eq:social_welfare2}
\begin{gathered}
\text { (II): } \max \sum_{\substack{i \in N \\
\varphi \in \Phi}} C_{i_{\varphi}}^c\left(P_{i_{\varphi}}^c\right)-C_{i_{\varphi}}^g\left(P_{i_{\varphi}}^g\right)-C_i^{VU}\left(f\left(v_i\right)\right) \\
\end{gathered}
\end{equation}
\begin{equation}
\hspace*{0mm}\text{subject to: \quad constraints~(\ref{eq:P_balance1}-\ref{eq:Thermal1})} \nonumber
\end{equation}

The objective of (\ref{eq:social_welfare2}) is to maximize social welfare by incorporating the cost associated with voltage unbalance on the grid. A sensitivity analysis of this cost component will be conducted in subsequent sections.

The Lagrangian function becomes:
\begin{equation}
\label{L_2}
\begin{aligned}
\mathcal{L} = & \sum_{\substack{i,j \in N \\ \{i,j\} \in L \\ {\varphi} \in \Phi}} \Bigg[ C_{i_{\varphi}}^g (P_{i_{\varphi}}^g) + C_i^{VU} (f(v_i)) - C_{i_{\varphi}}^c (P_{i_{\varphi}}^c) \\
& + ... + \eta_{i j, \varphi} (p_{i j, \varphi}^2 + q_{i j, \varphi}^2 - \bar{s}_{i j, \varphi}^2) \Bigg]
\end{aligned}
\end{equation}
The Lagrangian function in Eq. (\ref{L_2}) is identical to that in Eq. (\ref{L_1}), except that the last term of Eq. (\ref{L_1}) is replaced by the second term of Eq. (\ref{L_2}). The term  \(C_i^{VU} (f(v_i))\) represents the cost associated with voltage unbalance rather than a grid limit.
Subsequently, the new stationary condition with respect to $P_{n_{\varphi}}^c$ results in the DLMP of active power demand for node $n$ and on phase $\varphi$ to be equal to:
\begin{equation}
\label{Penal}
\begin{aligned}
& \text{DLMP}_{n,{\varphi}} = \\
& \textrm{c}^{\textrm{VU}}_{\textrm{i}} \frac{\partial f(v_n)}{\partial P_{n_{\varphi}}^c} + ... + 2 \sum_{\{i,j\} \in L} \eta_{ij,{\varphi}} \left( p_{ij_{\varphi}} \frac{\partial p_{ij_{\varphi}}}{\partial P_{n_{\varphi}}^c} + q_{ij_{\varphi}} \frac{\partial q_{ij_{\varphi}}}{\partial P_{n_{\varphi}}^c} \right)
\end{aligned}
\end{equation}
The term capturing the impact of the voltage unbalance penalty corresponds to the first term in Eq. (\ref{Penal}); the remaining components are similar to Eq. (\ref{Cons}) except for the omission of its last term. The function retains the structure introduced previously; therefore, its influential factors will be examined in Subsection \ref{math_math} to provide deeper insight into its characteristics. Detailed discussions on the decomposition of energy, losses, congestion, and voltage magnitude limits can be found in \cite{papavasiliou2017analysis, yuan2016novel}.

\subsection{Detailed Decomposition of Voltage Unbalance Components}
\label{math_math}
We now provide a detailed decomposition of the components contributing to voltage unbalance within the DLMP framework. The formulation applies to both \textit{soft limit} and \textit{hard limit} approaches. The distinction between these two methods lies in the multiplier associated with the term containing the partial derivative of the voltage unbalance metric with respect to $P_{n_{\varphi}}^c$: while this term is a constant in the \textit{soft limit} approach, it is  a dual variable in the \textit{hard limit} approach. Term $\frac{\partial f(v_n)}{\partial P_{n_{\varphi}}^c}$ is therefore expanded, for which the impact of $ P_{n_{\varphi}}^c $ on the voltage unbalance of nodes other than $n$ is neglected. 

The objective is to determine $\frac{\partial f(v_n)}{\partial P_{n_{\varphi}}^c}$, which represents the voltage unbalance changes in the node $n$, due to the small changes in the power demand in the same node and phase $\varphi$.
The first step considers a general three-phase system, where the chain rule is applied to derive the following expression:
\begin{equation}
\label{eq:prejacob}
\frac{\partial f(v_n)}{\partial P_{n}^c} = \frac{\partial f(v_n)}{\partial v_n}\cdot{\frac{\partial v_n}{\partial P_{n}^c}}
\end{equation}
Using the Jacobian matrix, equation (\ref{eq:prejacob}) can be rewritten as:
\begin{equation}
\label{eq:jacobian}
J_f(P_{n}^c) = J_f(v_n)J_v(P_{n}^c)
\end{equation}
It can be proven that small changes of voltages in two other phases are negligible, so only diagonal elements will appear, and (\ref{eq:jacobian}) for a single phase can be simplified as:
\begin{equation}
\label{eq:single_phase_jacob}
\frac{\partial f(v_n)}{\partial P_{n_{\varphi}}^c} = \frac{\nabla_{v_{n, \varphi}}f(v_n) \cdot \nabla_{v_{n, \varphi}}P_{n_{\varphi}}^c}{\Vert\nabla_{v_{n, \varphi}}P_{n_{\varphi}}^c\Vert^2}
\end{equation}
Equation (\ref{eq:single_phase_jacob}) provides the directional sensitivity of voltage unbalance index with respect to active power \cite{boyd-replacement}.

The most widely adopted definition of voltage unbalance is the Voltage Unbalance Factor (VUF), which quantifies the ratio of negative sequence voltage to positive sequence voltage in a three-phase system. This metric provides a standardized measure of unbalance severity. The formal definition of VUF is given in \cite{IEC} and \cite{IEEE1159}.
\begin{equation}
\begin{aligned}
\text{VUF} = \frac{\Vert v_a + a^2 v_b + a v_c \Vert}{\Vert v_a + a v_b + a^2 v_c \Vert}
\end{aligned}
\end{equation}
In which \( a = e^{j 120^\circ} \). 

It represents the ratio of the magnitudes of the negative and positive sequence components at a given node in a three-phase system. Since these components are expressed as phasors, the calculation involves taking the square root of the product of the phasor and its complex conjugate. To reduce computational complexity, the square root operation is relaxed, and $f(v_n)$ is defined as:
\begin{equation}
\begin{aligned}
f(v_n) &= \text{VUF}^2(v_n) \\ 
&= \frac{v_a + a^2 v_b + a v_c}{v_a + a v_b + a^2 v_c}\left[\frac{v_a + a^2 v_b + a v_c}{v_a + a v_b + a^2 v_c}\right]^*
\end{aligned}
\end{equation}

It should be noted that the node index is removed from the phase voltages for the sake of clarity and simplicity. In other words, because this index is defined exclusively in a node, $v_a$ is actually $v_{n,a}$.

So $\nabla_{v_{n, \varphi}}f(v_n)$ for a generic phase (without loss of generality, phase \textit{`a'} is used) can be derived using \textit{Matrix Wirtinger Calculus} \cite{Wirtinger} as follows:
\begin{equation}
\label{eq:VUF^2}
\begin{aligned}
&\nabla_{v_{n, a}}f(v_n) = \\
&\frac{j \sqrt{3} \;v_{bc} \left( v_a^{*2} + v_b^{*2} + v_c^{*2} - v_a^* v_b^* - v_b^* v_c^* - v_c^* v_a^* \right)}{D^2} \\
&- \frac{j \sqrt{3} \;v_{bc}^{*} \left( v_a^{2} + v_b^{2} + v_c^{2} - v_a v_b - v_b v_c - v_c v_a \right)}{D^2} \\
&= \frac{\sqrt{3} \;\Im\{v_{bc}^{*} (v_{ab}^{2} + v_{bc}^{2} + v_{ca}^{2})\}}{D^2}
\end{aligned}
\end{equation}
Where $D$ is the square of positive sequence voltage magnitude, defined as:
\begin{align}
D &= (v_a + a v_b + a^2 v_c)(v_a + a v_b + a^2 v_c)^* \nonumber\\
&= \Vert v_a + a v_b + a^2 v_c \Vert ^2
\end{align}

In the end, $\nabla_{v_{n, \varphi}}P_{n_{\varphi}}^c$ for a generic phase results in:
\begin{equation}
\begin{aligned}
\label{eq:current}
\nabla_{v_{n, a}}P_{n_{a}}^c = \sum_{j: \{n,j\} \in L} I_{nj,a}^*
\end{aligned}
\end{equation}

So the second part of (\ref{eq:single_phase_jacob}) would be:
\begin{equation}
\label{eq:current_last}
\begin{aligned}
\frac{\nabla_{v_{n, a}}P_{n_{a}}^c}{\Vert \nabla_{v_{n, a}}P_{n_{a}}^c \Vert^2} = \frac{\sum_{j: \{n,j\} \in L} I_{nj,a}^*}{\Vert\sum_{j: \{n,j\} \in L} I_{nj,a}^*\Vert^2}
\end{aligned}
\end{equation}

Finally, $\frac{\partial f(v_n)}{\partial P_{n_{\varphi}}^c}$ in node $n$ for a generic phase using  (\ref{eq:single_phase_jacob}), (\ref{eq:VUF^2}) and (\ref{eq:current_last}) will be:
\begin{equation}
\label{eq:complex_last}
\begin{aligned}
\frac{\partial f(v_n)}{\partial P_{n_a}^c} = \frac{\sqrt{3} \;\Im\{v_{bc}^{*} (v_{ab}^{2} + v_{bc}^{2} + v_{ca}^{2})\}}{D^2 \Vert\sum_{j: \{n,j\} \in L} I_{nj,a}^*\Vert^2} \cdot \nabla_{v_{n, a}}P_{n_{a}}^c
\end{aligned}
\end{equation}

Given the inherent complexity of~(\ref{eq:complex_last}), it becomes essential to extract meaningful interpretations that reveal its underlying structure. These interpretations not only serve as powerful analytical instruments, but also provide a foundation for deeper investigation and enhanced insight. For clarity, they are articulated with respect to the generic phase of \textit{a}; however, it should be emphasized that these statements hold universally across all phases.

\begin{enumerate}
  \item \label{opposite VL}
    In (\ref{eq:VUF^2}), the only asymmetric component is $v_{bc}^{*}$, which corresponds to the opposite line voltage of phase \textit{`a'}. This asymmetry indicates that the derivative at the same node behaves differently across phases, depending on the magnitude and interaction of the opposite line voltage. Such behavior may produce counterintuitive effects. In an unbalanced three-phase system, the phase with an intermediate voltage magnitude lies between those with the highest and lowest voltages. The opposite line-to-line voltage—formed between these two extreme phases—is typically the largest, as it represents the difference between the maximum and minimum phase voltages. Consequently, according to the given expression, the middle phase may experience a higher penalty than either the most or the least heavily loaded phase. This phenomenon is analyzed in detail in Sections~\ref{TC1} and \ref{TC2}.
  \item \label{line voltage balance}
    Any deviation from balanced line voltages at a bus increases overall voltage unbalance. In a perfectly balanced system, the second term in the numerator of equation (\ref{eq:VUF^2}), $(v_{ab}^{2} + v_{bc}^{2} + v_{ca}^{2})$, is equal to zero. When line voltages diverge from this condition, the magnitude of this term rises, thereby amplifying voltage unbalance. This effect interacts with the asymmetry discussed in the previous point, understanding this interaction is essential for interpreting the behavior of ~(\ref{eq:complex_last}).
  \item \label{main line}
    From the denominator of ~(\ref{eq:complex_last}), nodes that inject significant current into a given phase exhibit lower sensitivity of the DLMP voltage unbalance term to active power fluctuations within that same phase. This occurs because the derivative of the unbalance metric with respect to power is scaled by the local current contribution, reducing its marginal effect. Consequently, nodes closer to the main feeder—where currents are typically higher—tend to have smaller DLMP adjustments for voltage unbalance compared to nodes farther downstream.
\end{enumerate}
By leveraging these three points as analytical instruments, one can systematically interpret and substantiate the behavior and implications of voltage unbalance constraints, as well as the associated penalty terms.

It is important to highlight that the development of DLMP for reactive power is fundamentally similar to that for active power, the key distinction being the absence of an explicit cost for reactive power. So the $\frac{\partial f(v_n)}{\partial Q_{n_{\varphi}}^c}$ will be:
\begin{equation}
\begin{aligned}
\frac{\partial f(v_n)}{\partial Q_{n_a}^c} = \frac{\sqrt{3} \;\Im\{v_{bc}^{*} (v_{ab}^{2} + v_{bc}^{2} + v_{ca}^{2})\}}{D^2 \Vert\sum_{j: \{n,j\} \in L} I_{nj,a}^*\Vert^2} \cdot \nabla_{v_{n, a}}Q_{n_{a}}^c
\end{aligned}
\end{equation}

In summary, OPF formulations for hard and \textit{soft limit} approaches were developed, and DLMP decompositions derived. The voltage unbalance component was analyzed to identify key influencing factors, providing tools that improve DLMP interpretation and support the proposed methods.

\section{Test Cases and Analysis}
\label{sec:test}
In this section, two test cases are presented that aim to highlight different aspects of the methodology introduced:

\begin{itemize}
    \item \textbf{Test Case 1:} A simple and small-scale network is used to demonstrate the functionality and effectiveness of the proposed unbalance-aware OPF formulations. The primary objective is to compare the \textit{hard} and \textit{soft limit} methodologies and validate the analytical results from the previous section and to illustrate how they can be used to interpret and justify OPF behavior under unbalanced conditions.
    \item \textbf{Test Case 2:} The second test case is based on the IEEE standard European low-voltage (LV) distribution network. This scenario is designed to assess the scalability and practical applicability of the proposed methods, with particular emphasis on the \textit{soft limit} approach. The results highlight its improved performance relative to the conventional \textit{hard limit} method.
\end{itemize}

All test cases are performed using the \texttt{PowerModelDistribution} package~\cite{FOBES2020106664}. The voltage unbalance constraint and penalty factor are implemented externally in \texttt{Julia}~\cite{Bezanson2017} using \texttt{JuMP}~\cite{Dunning2017}, and integrated into the main OPF formulation. The solver used for all simulations is \texttt{Ipopt}~\cite{Ipopt2006}. For reproducibility, the solver configuration are: (tolerances = 1e-4, maximum iterations = 10000), and all simulations were performed on a laptop equipped with an Intel Core i7-1255U CPU @ 1.70 GHz, 16 GB of RAM, running Windows 11.
The developed code is publicly available on GitHub.\footnote{GitHub Repository: \url{https://github.com/alireza33zz/VU-DLMP_Framework-v1.0.git}}

\subsection{Test Case 1: Simple Network}
\label{TC1}
\begin{figure}[!t]
\centering
\includegraphics[width=1.0\linewidth]{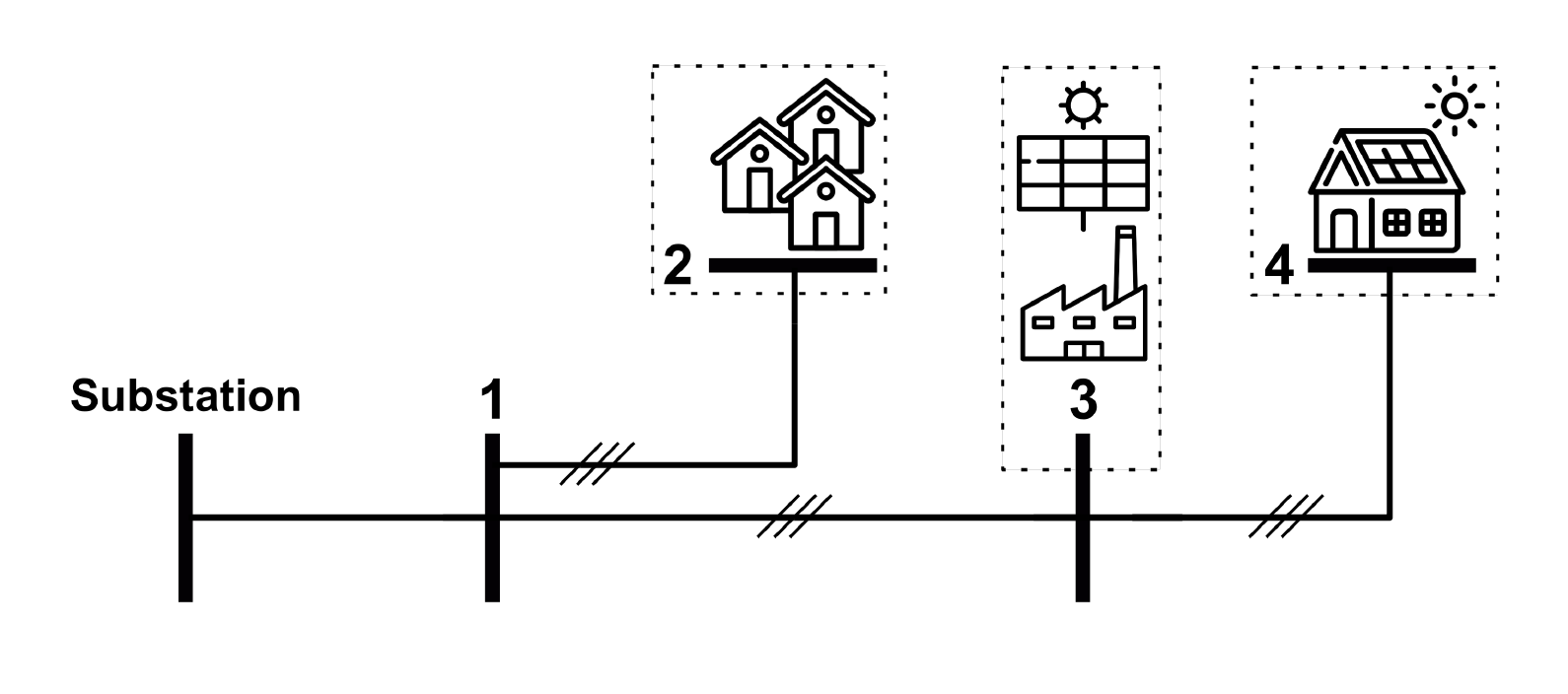}
\caption{Single-line diagram of test case 1, `simple network'.}
\label{fig:Schem_simpleGrid}
\end{figure}

The proposed network has 5 buses, connected to the grid through a substation and represents: 
\begin{itemize} 
\item Unbalanced loads on bus 2,
\item Balanced loads, and a 3-phase photovoltaic generator on bus 3,
\item Unbalanced loads with single-phase PV rooftop panels on bus 4.
\end{itemize}
A schematic of this network is presented in Fig.~\ref{fig:Schem_simpleGrid}.

\begin{figure}[!t]
    \centering
    \includegraphics[width=1.0\linewidth]{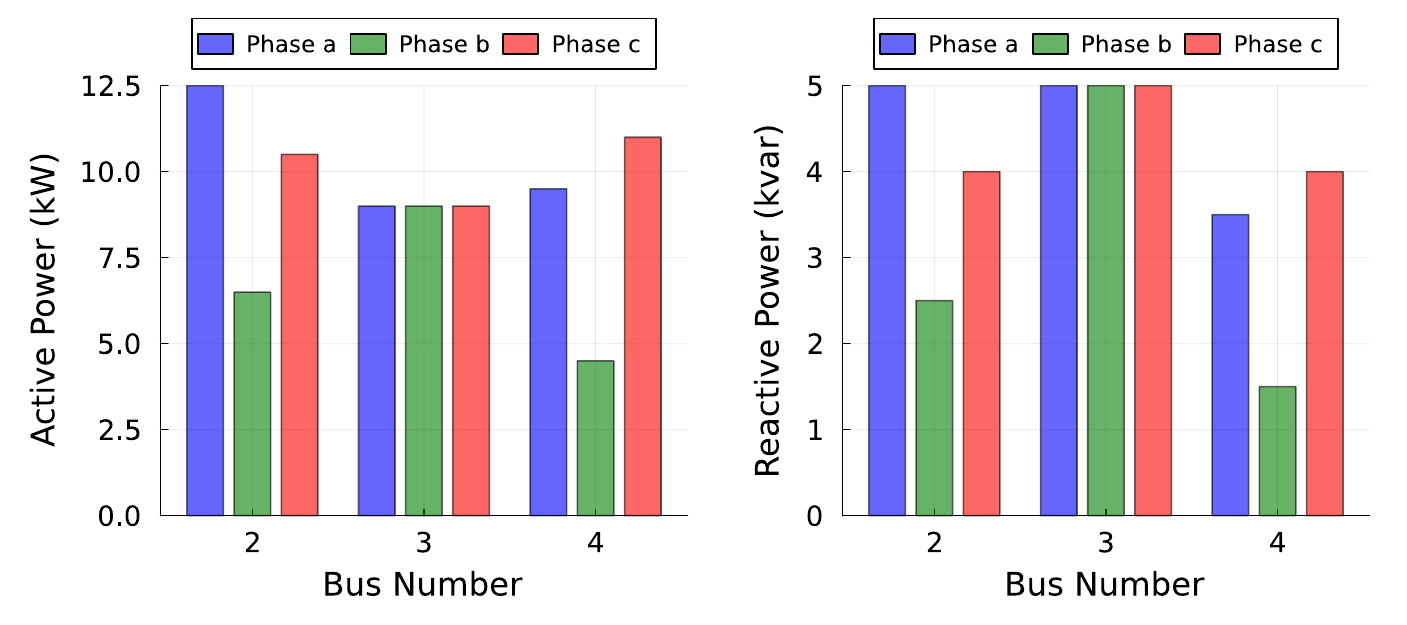}
    \caption{Active and reactive loads of `simple network'.}
    \label{fig:active_loads_simple}
\end{figure}

The sizes of active and reactive loads connected to each phase and node are shown in Fig.~\ref{fig:active_loads_simple}. The total active power demand in this network is 81.5 kW. The distribution of loads between phases is: 38.5\% for phase $a$, 24.5\% for phase $b$ and 37\% for phase $c$. 

The substation marginal energy cost is set to $1\,\text{€/kWh}$ as a normalization reference for reporting DLMP components; scaling this value primarily rescales prices in €/kWh without changing the qualitative comparisons emphasized in this study , while the cost for renewable generators is set to 0 €/kWh.
Two types of distributed generation are considered in the system:

a) Bus 3 hosts a three-phase balanced photovoltaic (PV) generator with a nominal capacity of 21 kVA. This generator is capable of both injecting and absorbing reactive power up to $\pm$18 kvar.

b) Bus 4 includes three single-phase rooftop PV panels, each with a nominal capacity of 2.5 kVA, operating at unity power factor \cite{rodriguez2021probabilistic}.

\begin{figure*}[!t]
\centering
    \includegraphics[width=0.80\linewidth]{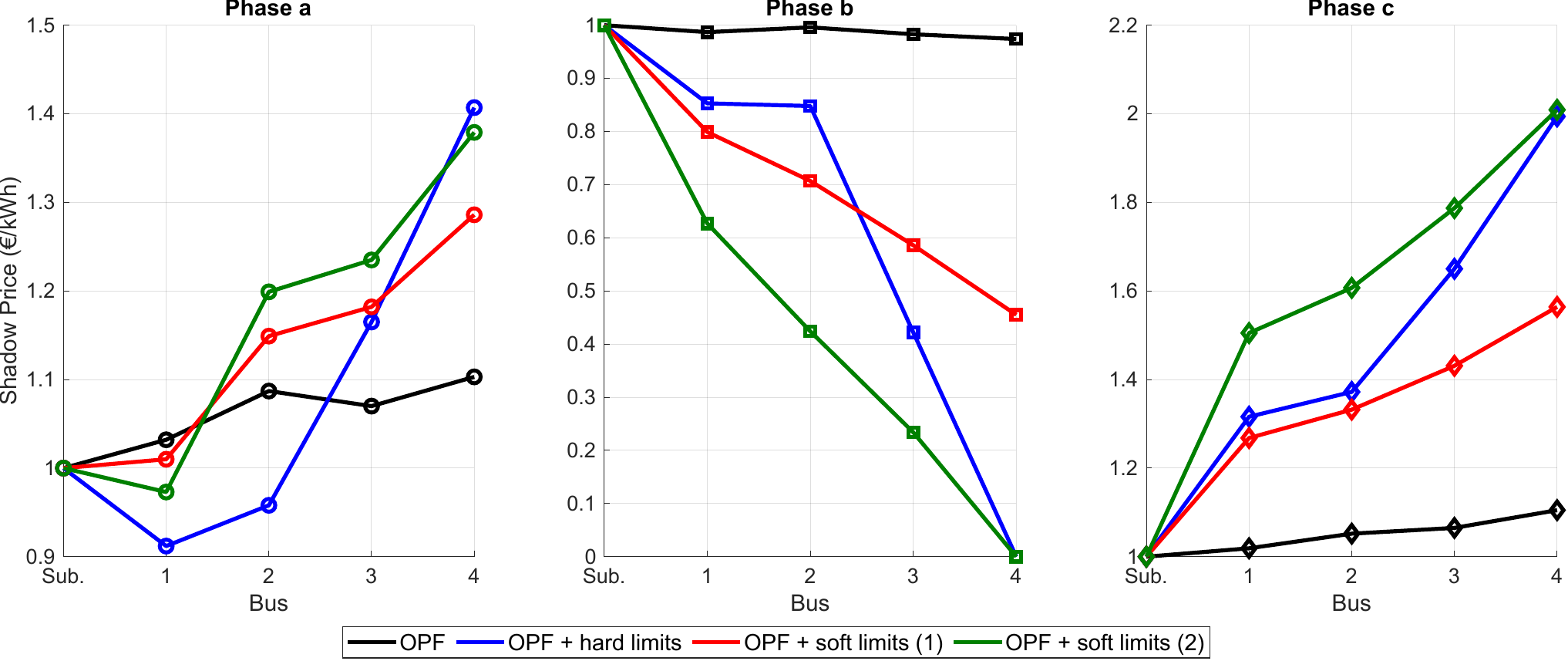}
\caption{Shadow prices of active power of `simple network' under four scenarios.}
\label{fig:Shadow_simpleGrid_P}
\end{figure*}

\begin{figure*}[!t]
\centering
    \includegraphics[width=0.80\linewidth]{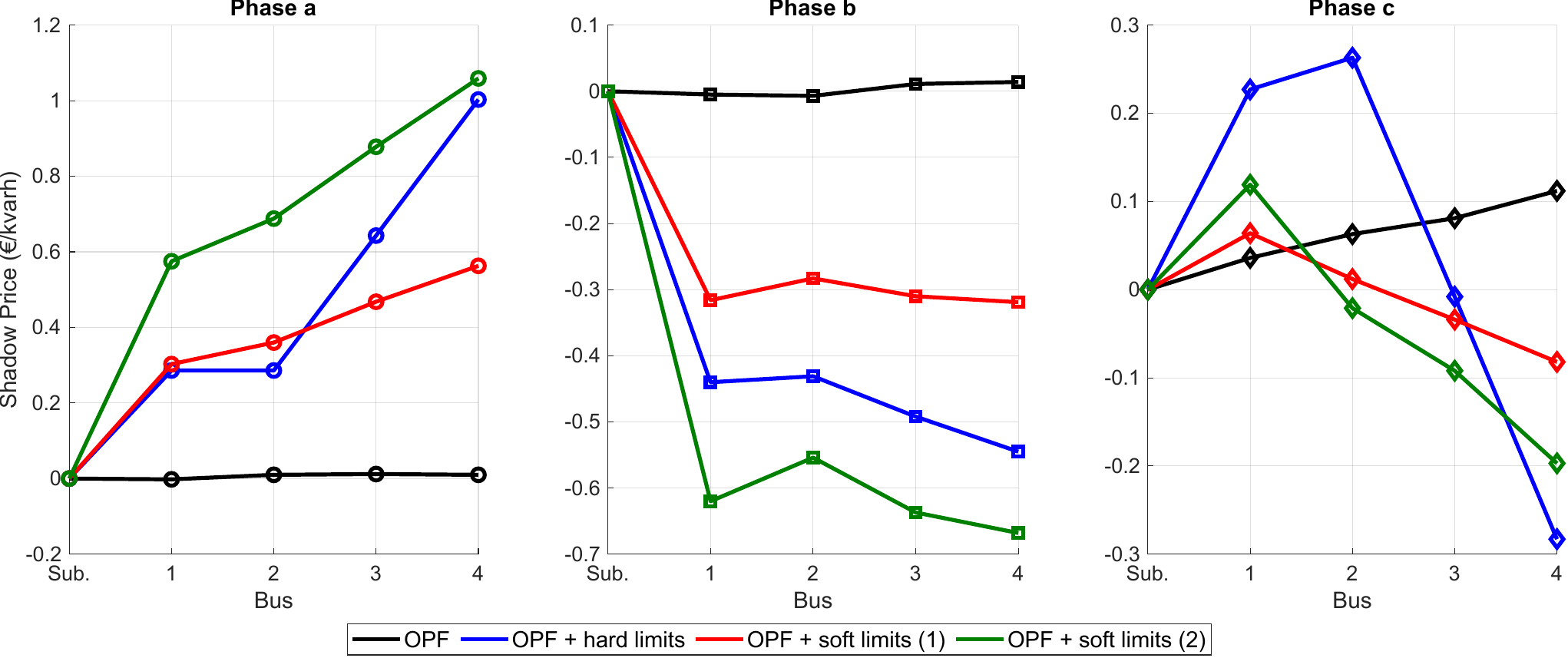}
\caption{Shadow prices of reactive power of `simple network' under four scenarios.}
\label{fig:Shadow_simpleGrid_Q}
\end{figure*}

Four different case scenarios for this network are presented. All simulations are performed over a 1-hour time interval, i.e., corresponding to a single run of the three-phase OPF. The total generation cost, total losses, and the highest percentage of VUF in a node in the four different scenarios are presented in Table~\ref{table:test_case_results}. 
Four cases are considered: (1) baseline AC-OPF (no VU integration), (2) AC-OPF with hard VU limits, (3.a) and (3.b) AC-OPF with soft VU limits using two different coefficients.
The shadow prices of these scenarios are illustrated in Figs.~\ref{fig:Shadow_simpleGrid_P} and~\ref{fig:Shadow_simpleGrid_Q}. The four cases are as follow:
\begin{table}[!t]
    \captionsetup{justification=centering, textfont={sc,footnotesize}, labelfont=footnotesize, labelsep=newline} 
    \renewcommand{\arraystretch}{1.2}
\centering
\caption{Test case results for `simple network'.}
\begin{tabular}{>{\centering\arraybackslash}p{1.5cm}>{\centering\arraybackslash}p{1.5cm}>{\centering\arraybackslash}p{1.8cm}>{\centering\arraybackslash}p{1.5cm}}
\toprule
{\footnotesize \textbf{Test case number}} & 
{\footnotesize \textbf{Total gen. cost (€)}} & 
{\footnotesize \textbf{Total losses (kWh})} & 
{\footnotesize \textbf{Highest VUF (\%)}} \\
\midrule
1 & 55.21 & 2.20 & 1.11 \\
2 & 56.08 & 2.18 & 1.00 \\
3.a & 55.21 & 2.20 & 1.11 \\
3.b & 56.86 & 2.17 & 0.90 \\
\bottomrule
\end{tabular}
\label{table:test_case_results}
\end{table}

\textbf{1. Default OPF:} In this scenario, the nodal shadow prices are obtained from a fuel-cost minimization problem. In this formulation, fuel-cost minimization is equivalent to social welfare maximization because demand is assumed to be inelastic and consumer utility remains constant. This case facilitates the identification of general DLMP trends resulting from network losses, load imbalance, and operational constraints, including voltage limits.

\textbf{2. OPF with voltage unbalance as constraint:} In this scenario, the nodal shadow prices in a normal fuel cost minimization problem while imposing a voltage unbalance threshold of 1\% are obtained. It not only highlights the impact of previously discussed factors, but also illustrates the influence of the VUF constraint. 
Increasing the active load in phase $b$, which has the least load, incurs a lower cost compared to the marginal cost of generation at the substation bus. This effect is attributed to the associated reduction in voltage and phase angle deviation. Furthermore, adding reactive load in this phase can lead to negative prices, indicating potential revenue generation. This occurs because the reduction in voltage magnitude improves voltage unbalance.
Phase $c$, being moderately loaded, experiences the highest opposite line voltage, resulting in a significant excessive cost for its active power. This phenomenon was previously explored in item~\ref{opposite VL} of interpretations. Interestingly, this phase can gain revenue at bus 4 by adding reactive power, which causes a voltage drop that reduces voltage unbalance at that bus.
Phase $a$, which has the highest load, faces higher prices for both active and reactive power. The excessive cost for reactive power is particularly severe, being twice that of active power, due to the critical voltage drop situation in this phase.

If no constraint is binding—meaning all buses exhibit a VUF value below the specified threshold (e.g., 2\%)—the results remain unchanged compared to Case~1. In such a scenario, no economic signals related to voltage unbalance are generated.

\textbf{3. OPF with voltage unbalance penalization:} In this scenario, the nodal shadow prices obtained from an OPF formulation incorporating voltage unbalance penalization are presented. Two weighting factors are considered to evaluate the sensitivity of the results to the magnitude of penalties. This sensitivity analysis is relevant because the penalty reflects the economic impact of voltage unbalance rather than imposing \textit{hard limits}. Different customers may tolerate varying degrees of unbalance depending on their assets, which suggests that penalties could eventually be node-specific. Although this work does not explore node-specific penalties, the analysis illustrates how adjusting the penalty influences the social optimum by internalizing the negative effects of voltage unbalance.

\textbf{a.} In the first case, it is assumed that the penalization value of voltage unbalance on all nodes is equal to the marginal cost of generation in the substation, which means that it is equal to 1 €/h.\%.

\textbf{b.} In the second case, it is assumed that the penalization value of voltage unbalance on all nodes is equal to 1.5 times more than the marginal cost of generation in the substation, which means that it is equal to 1.5 €/h.\%. 

The results of both the 3.a and 3.b cases are presented in Figs.~\ref{fig:Shadow_simpleGrid_P} and~\ref{fig:Shadow_simpleGrid_Q}.    
These cases exhibits similar trends in the DLMP of active and reactive power as observed in Case 2, across both weighting scenarios. Although the specific values differ due to the different tuning in place, the trends and relative comparisons remain consistent.
When a weighting ratio of 1:1.5 is applied between generation cost and the voltage unbalance penalty, the distinctions become more pronounced compared to a 1:1 ratio. The primary differences appear in power flow and dispatch decisions, which will be discussed in detail later, and in their effectiveness in reducing voltage unbalance levels, as shown in Table~\ref{table:test_case_results}. The choice of penalty value is critical because it should reflect the economic impact of voltage unbalance on network assets. For example, higher penalties may be justified when sensitive equipment or customer-owned assets are more susceptible to accelerated wear and tear under unbalanced conditions. This approach provides a more realistic representation than \textit{hard limits}, as it internalizes the cost of degradation and enables operators to balance system efficiency with asset protection.
In Case~3.a, the dispatch remains unchanged from Case~1 because the penalization weight is insufficient to influence generation decisions. Nevertheless, the internalized cost of voltage unbalance is reflected in DLMP values, even without changes in dispatch. In contrast, Case~3.b achieves a VUF reduction below 1\%, which is assumed to comply with grid code requirements. In both cases, generation cost increases due to partial or complete curtailment of single-phase generator output, while grid losses exhibit a slight decrease. 

A key distinction is that the true cost of voltage unbalance is consistently represented in shadow prices in this formulation, unlike Case~2, where such signals appear only when the constraint becomes binding.

Overall, it is evident that nodes closer to the main feeder (substation) or situated on the primary power transfer lines are less sensitive to voltage unbalance. This observation was previously discussed in item~\ref{main line} of interpretations.

Negative DLMP values occur when an incremental increase in consumption reduces the system objective (e.g., by alleviating phase imbalance or reducing losses or constraint costs), hence reflecting a marginal system benefit rather than a subsidy. Settlement remains standard nodal pricing: the cleared quantity is settled at the nodal price, so a negative DLMP implies the consumer is compensated because the action is beneficial to system operation. Such outcomes are not unprecedented, as negative nodal prices can also occur in existing electricity markets when they reflect cost-minimizing operation under network and resource constraints.

As shown in the results, both VU integration approaches produce similar overall trends, indicating broadly comparable economic signals when VU constraints are non-binding. The differences arise from the distinct mechanisms of the two formulations. Under \emph{hard limits}, VU has no marginal impact as long as the corresponding constraint is inactive; however, once the constraint becomes binding (e.g., at bus~4), it directly reshapes the local feasible set and triggers a redispatch that modifies flows and voltages along the affected branch. Due to electrical coupling, this local change propagates through the network, with stronger effects on electrically adjacent buses and weaker effects on parallel sections (e.g., buses~1--2 remain similar, whereas buses~3--4 exhibit larger deviations). In contrast, the \emph{soft limit} formulation yields smoother behaviour because the VU penalty contributes to the objective at all operating points, so each bus can affect the optimal solution even when no explicit constraint is binding. Consequently, the soft limit method produces more coherent, continuous responses across the feeder, avoiding abrupt shifts induced by constraint activation and consistently reflecting VU impacts in the resulting DLMPs.

\subsection{Test Case 2: European LV Network}
\label{TC2}
\begin{figure}[!t]
    \centering
    \includegraphics[width=1.0\linewidth]{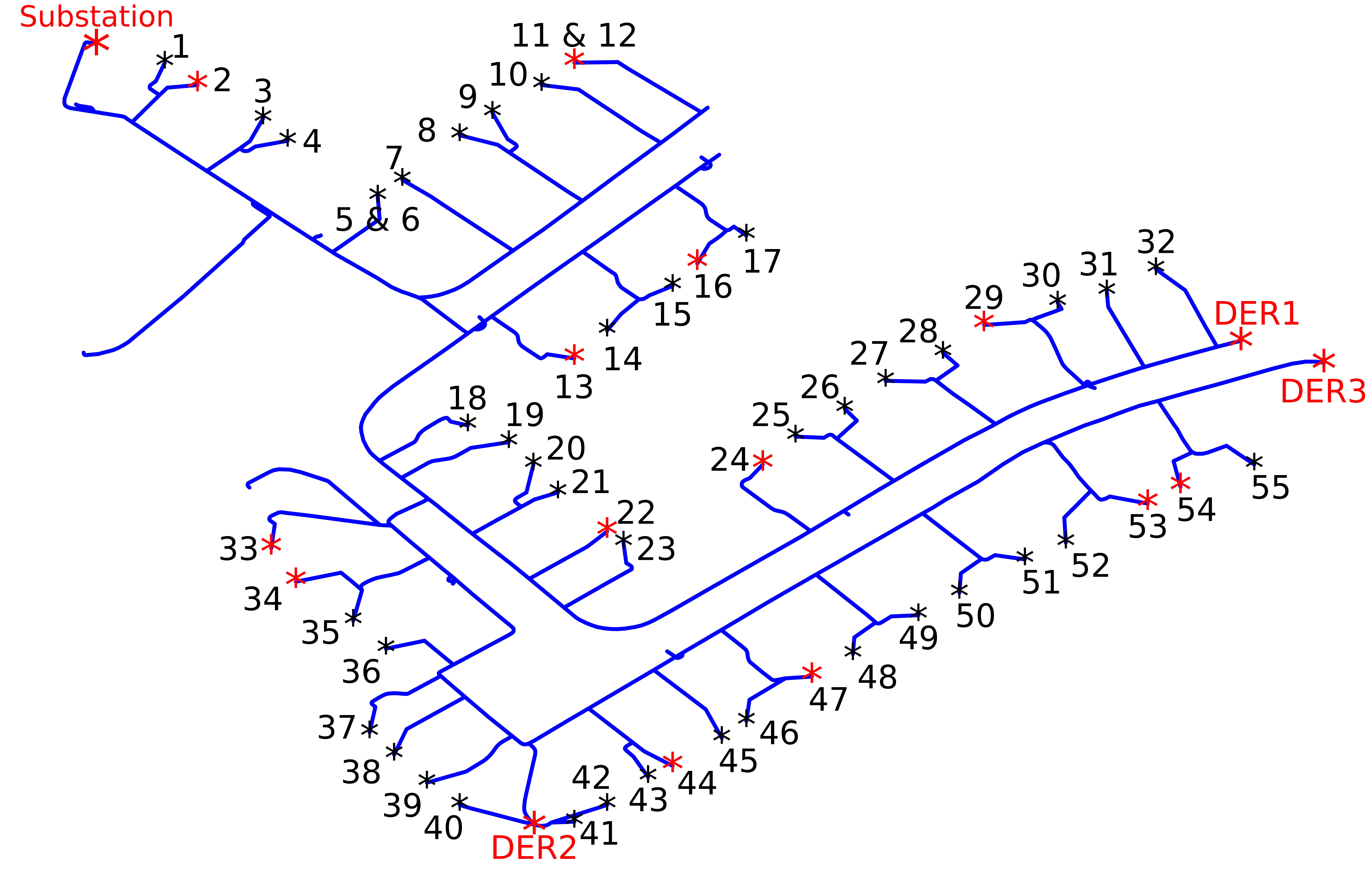}
    \caption{Schematic for the modified European LV network.}
    \label{fig:EU_LV_network_Schematics}
\end{figure}

The European low-voltage network consists of 55 three-phase load buses and a substation feeder~\cite{ieeeEuLV}. A modified version of this network is illustrated in Fig.~\ref{fig:EU_LV_network_Schematics}. The marginal cost of generation at the substation is fixed at 1\,€/kWh.

Three DER with reactive power support capabilities are integrated into the grid (indicated by large red stars in the figure):

\begin{itemize}
    \item DER2 is a solar power plant with a zero marginal generation cost and a capacity of 54\,kVA, and can provide up to 30\,kvar of reactive power support.
    \item DER1 and DER3 are inverter-based battery energy storage systems, each rated at 60\,kVA with a marginal dispatch cost of 1.1\,€/kWh (reflecting storage cycling cost \cite{Battery}). Both have a startup (activation) cost of 150\,€, and can provide up to 54\,kvar of reactive power support through their smart inverters \cite{Battery2}.
\end{itemize}

Additionally, 14 buses host rooftop PV panel arrays (marked with red stars), each with a capacity of 7.5\,kVA, operating at unity power factor. The total capacity of renewable generation is equal to 159 kW.

Three three-phase loads are connected to buses 9, 23, and 40, representing voltage unbalance-sensitive equipment, such as electric motors. Each has a demand of 18\,kW with a lagging power factor of 0.88. The remaining load buses are equipped with single-phase loads ranging from 3.5 to 5.5\,kW, operating at an average power factor of 0.93.

The total active power demand in the network is 287.5\,kW. The phase-wise load distribution is as follows:
\begin{itemize}
    \item Phase $a$: 38.3\%
    \item Phase $b$: 32.9\%
    \item Phase $c$: 28.8\%
\end{itemize}

The simulation setup follows the same configuration as the previous network. Although the official European LV benchmark model comprises 907 nodes, its large scale results in excessive computation times and convergence issues with commercial solvers. To overcome these limitations, a reduced-order model from~\cite{khan2022reduced} is adopted. This version preserves the exact electrical characteristics of the original network while reducing its size to 117 nodes, thereby maintaining full modeling fidelity with significantly lower computational complexity.

The comprehensive enforcement of VUF constraints across the entire network using the \textit{hard limit} method introduces significant computational challenges, often resulting in solver non-convergence. To mitigate this issue, the constraints are applied to a strategically selected subset of 15 critical buses. This subset includes voltage unbalance-sensitive loads, backbone buses, and nodes that violate VUF thresholds under standard OPF conditions. This targeted approach ensures protection of critical system components while avoiding excessive computational burden that would compromise solver stability.
The observed non-convergence under the full constraint set is mainly due to structural properties of the resulting nonlinear program rather than computational limitations. Additional constraints may induce degeneracy or near linear dependence, leading to an ill-conditioned KKT system and unreliable Newton directions \cite{NumOpt2006}. Furthermore, a significantly tightened or nearly infeasible feasible region degrades the effectiveness of interior-point methods that rely on strictly interior iterates \cite{Ipopt2006}. Differences in constraint scaling and the activation of highly restrictive constraints can further deteriorate numerical conditioning and prevent convergence \cite{NumOpt2006}.

\begin{table}[!t]
    \captionsetup{justification=centering, textfont={sc,footnotesize}, labelfont=footnotesize, labelsep=newline} 
    \renewcommand{\arraystretch}{1.2}
\centering
\caption{Comparison of convergence time of methods.}
\begin{tabular}{lcc}
\toprule
\textbf{\footnotesize Method} &
\makecell{\footnotesize \textbf{Number}\\\footnotesize \textbf{of nodes}} & 
\makecell{\footnotesize \textbf{Convergence}\\\footnotesize \textbf{Time (sec)}} \\
\midrule
Hard limits              & 56 & NaN \\
Soft limits (a)              & 56 & 19.81 \\
Soft limits (b)              & 56 & 22.78 \\
Relaxed hard limits      & 15 & 7.09 \\
Relaxed soft limits (a)      & 15 & 8.52 \\
Relaxed soft limits (b)      & 15 & 9.93 \\
\bottomrule
\end{tabular}
\label{tab:method_comparison}
\end{table}

For methodological consistency, the same subset is used in both the \textit{hard limit} and \textit{soft limit} formulations. Although the \textit{soft limit} method demonstrates robust numerical stability and can converge even when applied to all nodes, the selective subset is maintained to preserve analytical uniformity across scenarios, the convergence times of the different methods, considering both node sets, are reported in Table~\ref{tab:method_comparison}. As shown there, enforcing constraints on all nodes leads to non-convergence, whereas the soft limit method consistently converges. This superior numerical behavior of the \textit{soft limit} approach suggests that future work could exploit its scalability for full network implementations, enabling practical deployment in large scale systems.
Regarding scalability, it is acknowledged that enforcing VUF constraints for large networks remains computationally demanding \cite{churkin2024quantifying}. However, the proposed DLMP-based framework provides valuable insights into the economic signals associated with voltage unbalance, which is a novel contribution. 

\begin{figure}[!t]
    \centering
    \includegraphics[width=1.0\linewidth]{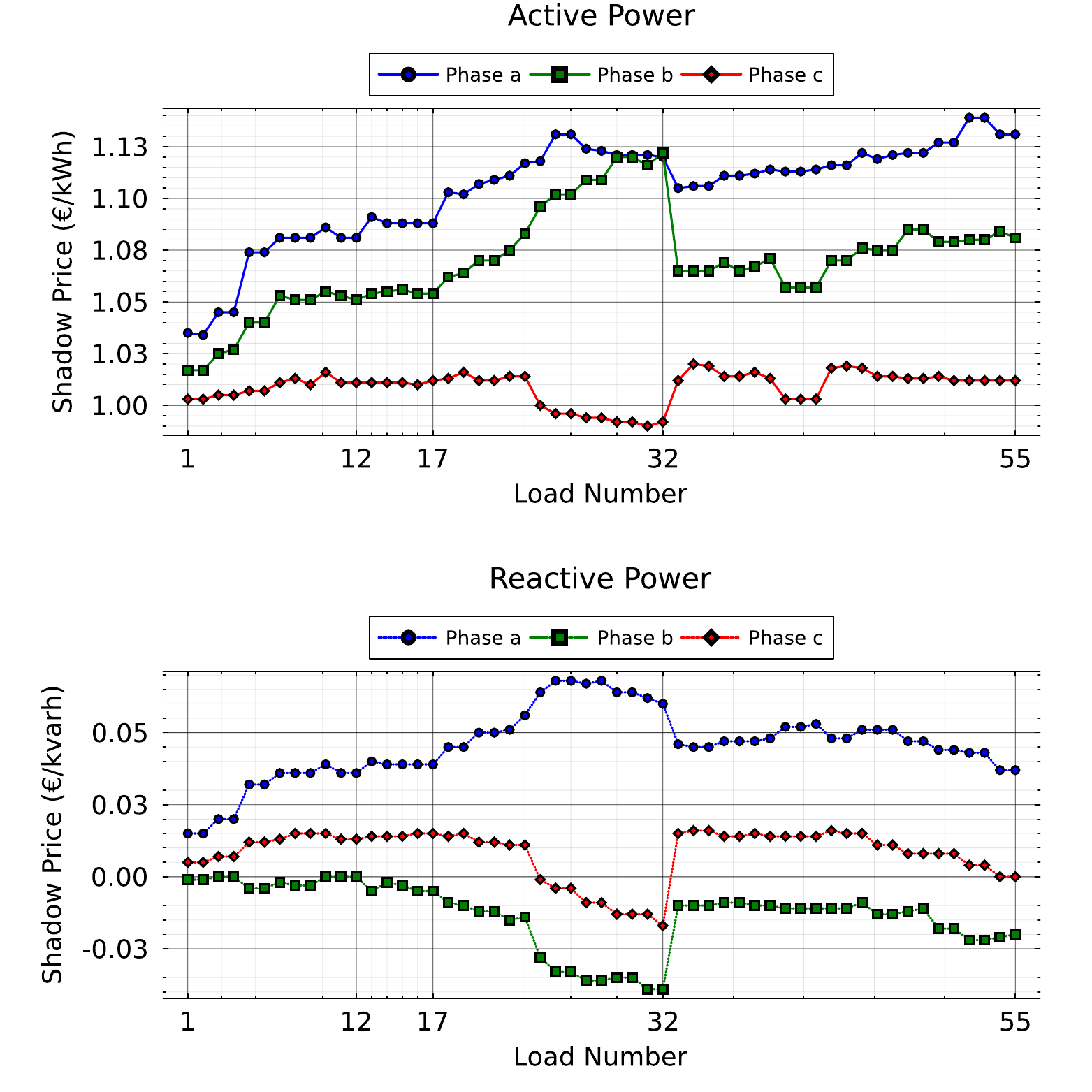}
    \caption{Shadow prices of OPF without VUF involvement.}
    \label{fig:combined_shadow_prices_default_case20}
\end{figure}

\begin{figure}[!t]
    \centering
    \includegraphics[width=1.0\linewidth]{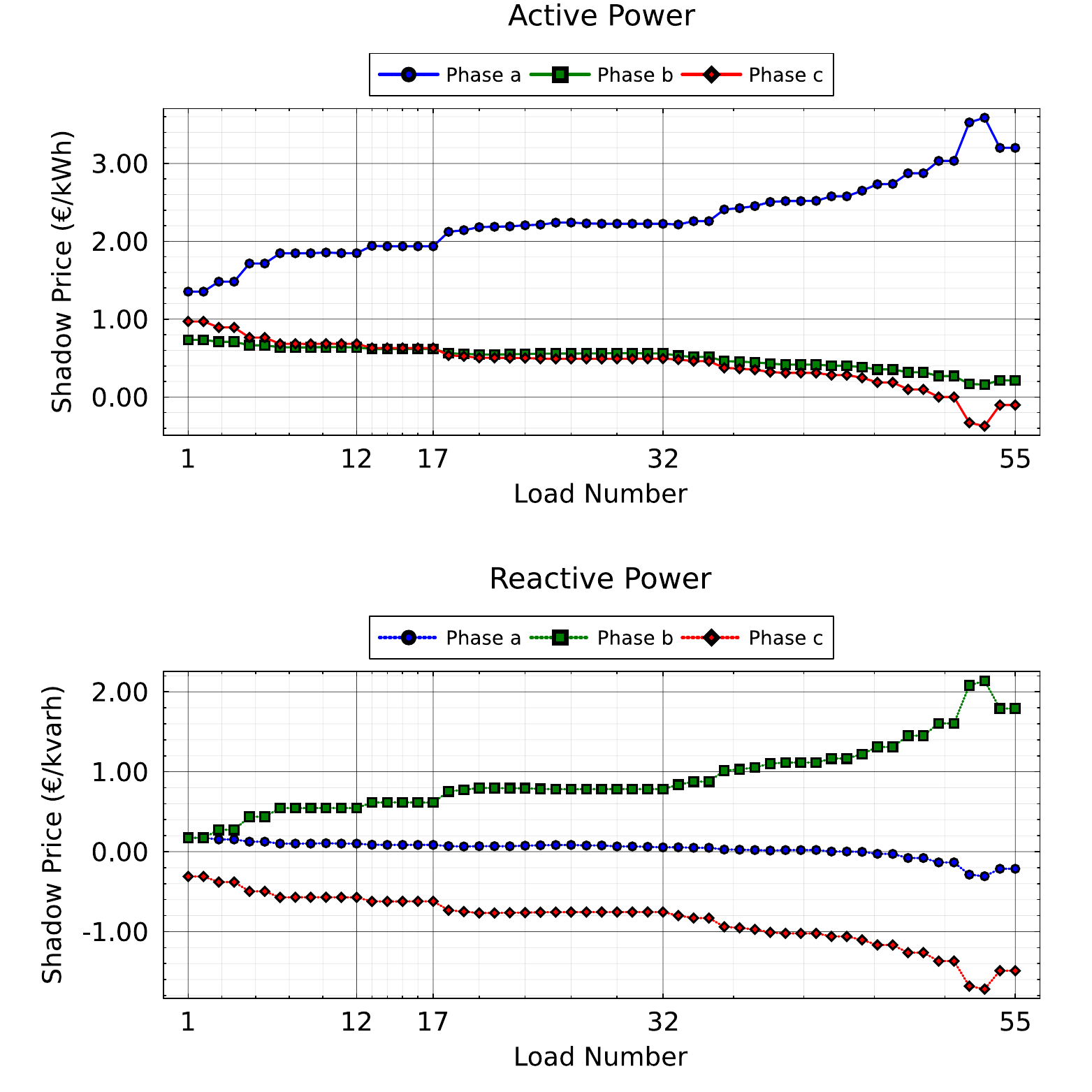}
    \caption{Shadow prices of OPF with VUF constraint.}    \label{fig:combined_shadow_prices_constraint_case20}
\end{figure}

\begin{figure}[!t]
    \centering
    \includegraphics[width=1.0\linewidth]{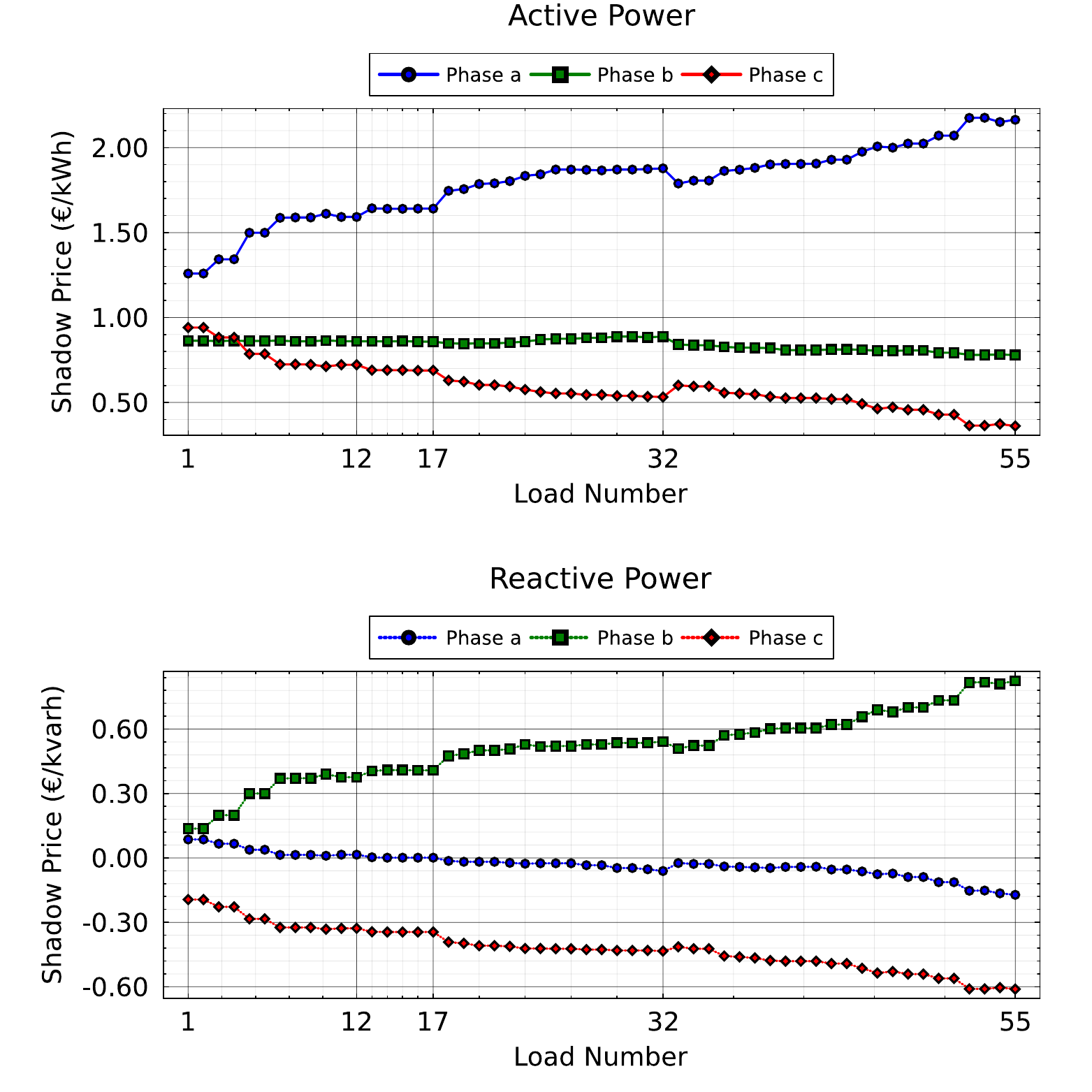}
    \caption{Shadow prices of OPF with weight of 1 to 1 VUF penalization (energy cost + VUF penalty).}    \label{fig:separate_shadow_prices_VUF+Gen_1.0_1.0_case20}
\end{figure}

\begin{figure}[!t]
    \centering
    \includegraphics[width=1.0\linewidth]{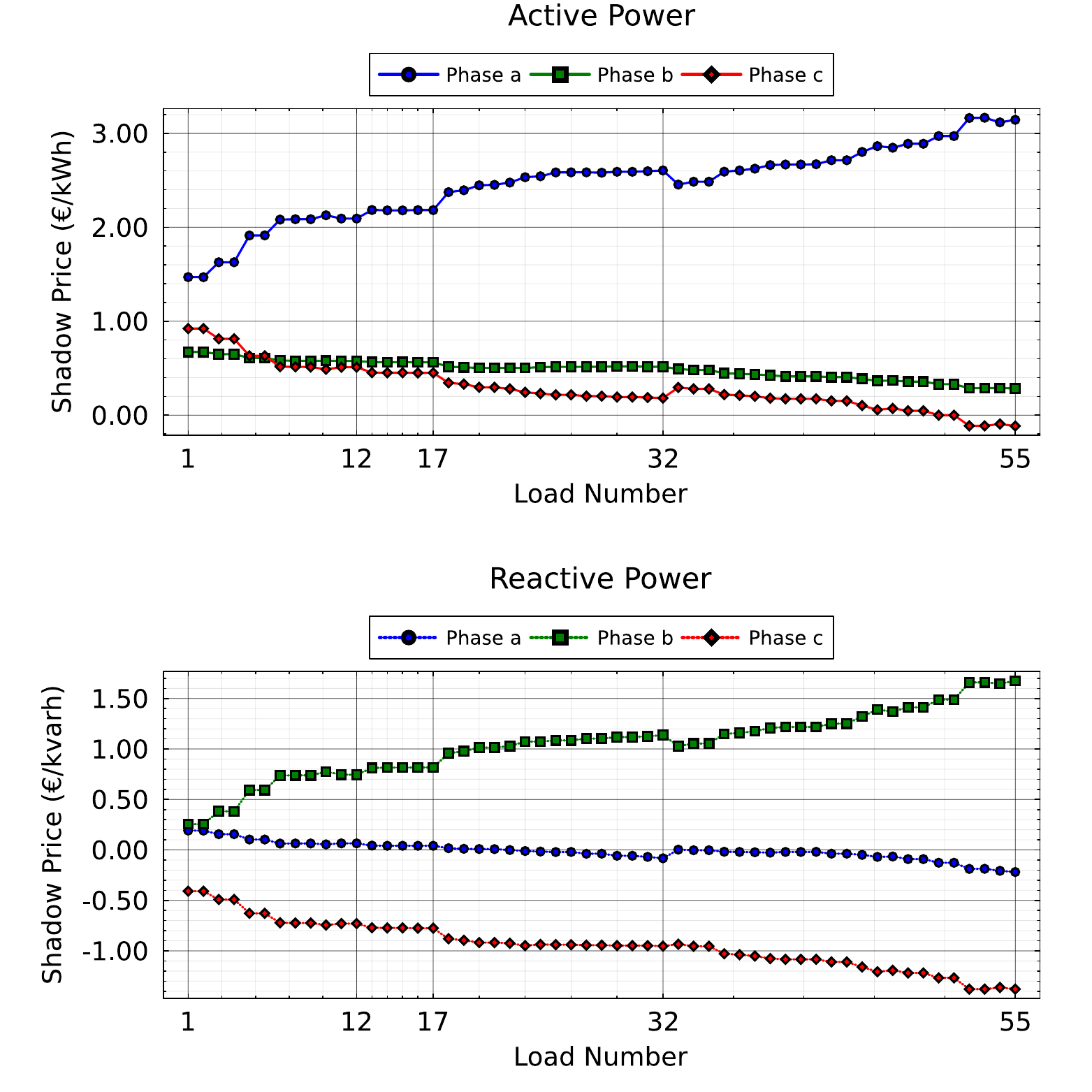}
    \caption{Shadow prices of OPF with weight of 1 to 1.5 VUF penalization (energy cost + 1.5 * VUF penalty).}    \label{fig:separate_shadow_prices_VUF+Gen_1.0_1.5_case20}
\end{figure}

The results of active and reactive power shadow prices for different scenarios 1, 2, 3.a and 3.b are presented in Fig.~\ref{fig:combined_shadow_prices_default_case20}, Fig.~\ref{fig:combined_shadow_prices_constraint_case20}, Fig.~\ref{fig:separate_shadow_prices_VUF+Gen_1.0_1.0_case20}, and Fig.~\ref{fig:separate_shadow_prices_VUF+Gen_1.0_1.5_case20} respectively.
The total generation cost, total losses, and the highest percentage of VUF in a node in the four different scenarios presented in Table~\ref{table:test_case_results2}.

\begin{table}[!t]
    \captionsetup{justification=centering, textfont={sc,footnotesize}, labelfont=footnotesize, labelsep=newline} 
    \renewcommand{\arraystretch}{1.2}
\centering
\caption{Test case results for `European LV network'.}
\begin{tabular}{>{\centering\arraybackslash}p{1.5cm}>{\centering\arraybackslash}p{1.5cm}>{\centering\arraybackslash}p{1.8cm}>{\centering\arraybackslash}p{1.5cm}}
\toprule
{\footnotesize \textbf{Test case number}} & 
{\footnotesize \textbf{Total gen. cost (€)}} & 
{\footnotesize \textbf{Total losses (kWh})} & 
{\footnotesize \textbf{Highest VUF (\%)}} \\
\midrule
1 & 434.7 & 6.15 & 1.13 \\
2 & 437.8 & 4.55 & 1.00 \\
3.a & 434.9 & 5.67 & 1.10 \\
3.b & 438.1 & 4.49 & 1.00 \\
\bottomrule
\end{tabular}
\label{table:test_case_results2}
\end{table}

Fig.~\ref{fig:combined_shadow_prices_default_case20} emphasizes the impact of losses, load imbalance between phases, and proximity to the main feeder.
Fig.~\ref{fig:combined_shadow_prices_constraint_case20} shows the impact of the voltage unbalance constraint. The shadow prices of active power of phase $a$ is very high due to the voltage drop from loading. The shadow prices of reactive power of phase $b$ is also very high due to the impact of the opposite line voltage. Adding load to phase $c$ can yield a lower marginal cost than the baseline energy component and, in some operating points, even negative DLMPs. This occurs because phase $c$ is the least loaded phase, so incremental consumption may (i) incur smaller marginal losses and congestion impacts and (ii) reduce binding penalty terms by improving the phase balance, thereby lowering the system-wide marginal cost. It is noteworthy to mention that total transmission losses have reduced significantly.

Fig.~\ref{fig:separate_shadow_prices_VUF+Gen_1.0_1.0_case20} and Fig.~\ref{fig:separate_shadow_prices_VUF+Gen_1.0_1.5_case20} exhibit similar trends to Fig.~\ref{fig:combined_shadow_prices_constraint_case20}, with both cases showing reduced losses. Case~3.a does not significantly reduce the VUF level because the penalization weight is relatively low, reflecting a scenario where customers are willing to tolerate higher unbalance to avoid additional costs. This outcome is not problematic; rather, it illustrates the flexibility of the penalization approach to accommodate different customer preferences. Conversely, Case~3.b achieves VUF reduction below 1\% with only a marginal increase in the generation cost, demonstrating how stronger penalization can improve voltage balance without compromising economic efficiency.

Across all three VU integration cases, the active power shadow prices follow intuitive loading patterns: the most heavily loaded phase exhibits the highest marginal cost, the least loaded phase the lowest, and the intermediate phase lies in between. In contrast, reactive power shadow prices are more sensitive to voltage coupling effects, particularly the influence of the opposite line voltage. As discussed in items~\ref{opposite VL} and~\ref{line voltage balance}, the observed phase behaviour is governed by the interaction between (i) phase loading and (ii) opposite line voltage, and the dominant factor depends on the load type and its spatial distribution. In \textbf{Test Case~2}, where three-phase loads deviate by 5\% from balance, the loading effect dominates the active power prices, while reactive power prices remain strongly shaped by voltage coupling.

These patterns imply that phase-targeted flexibility (e.g., shifting EV charging toward the least-loaded phase or PV panels curtailment) can reduce both operational costs and power quality penalties, and the corresponding DLMP components quantify this value transparently.

\subsection{Key Insights}
The main conclusions of these two test networks can be summarized as follows:

\begin{itemize}
\item The phase subject to the highest excessive cost is not necessarily the one causing the voltage unbalance. In many cases, the middle phase experiences significant excessive cost. However, when unbalance is primarily driven by voltage drop, the most heavily loaded phase is expected to incur the greatest excessive cost. This reflects the combined influence of these two factors (see items~\ref{opposite VL} and~\ref{line voltage balance} in the interpretation section).
\item Increasing loads can become economically advantageous compared to the base cost of energy at the substation. In certain phases, this may even result in negative prices effectively generating revenue when additional load reduces voltage unbalance.
\item Proximity to the main feeder enhances resilience against voltage unbalance impacts (see item~\ref{main line}).
\item Mitigating voltage unbalance generally leads to a reduction in network losses.
\end{itemize}

To summarize the performance of the \textit{soft limit} approach, its main advantage lies in the way it incorporates the economic impact of voltage unbalance directly into the objective function, rather than relying on threshold-based constraints as in the \textit{hard limit} method. This distinction is not merely computational; it reflects two fundamentally different operational philosophies. The \textit{hard limit} approach enforces strict compliance with technical standards, whereas the \textit{soft limit} approach internalizes the cost of unbalance, enabling trade-offs between technical performance and economic efficiency. Cases~3.a and~3.b illustrate how varying penalty weights influence these trade-offs, producing different DLMP patterns and cost signals that could guide operational practices and investment decisions. While the \textit{soft limit} formulation cannot guarantee strict adherence to VUF limits, it provides a flexible mechanism for signaling the economic consequences of voltage unbalance, which is the central focus of this analysis. Additionally, the \textit{soft limit} method exhibits significantly better numerical behavior and converges even when applied to all nodes, suggesting strong potential for addressing scalability challenges in future works. Future research could explore how these signals interact with asset degradation models and customer preferences to inform socially optimal solutions.

\subsection{Generality of results}
The qualitative conclusions of this work are expected to hold beyond the two test cases, but the magnitude and spatial distribution of the resulting phase-level DLMP components (including the unbalance related term) depend on several system specific factors. First, relative phase loading is a primary driver: larger phase asymmetries typically increase voltage and loss sensitivities and therefore amplify phase to phase price separation. Second, the location and type of unbalanced injections determine where VU is created or mitigated, shifting which buses/phases carry the largest marginal impacts. Third, the electrical distance to the substation shapes how strongly a local perturbation affects upstream power flow, voltages, and losses; this effect is moderated when parallel paths exist, which can redistribute flows and reduce local sensitivities. Finally, the numerical level of the unbalance related price component is sensitive to the adopted \emph{penalty parametrization}, which encodes the assumed economic valuation of power quality degradation, the impact of this factor is discussed in next section.

\subsection{Sensitivity analysis of VU marginal cost}
The VU penalty term ($C_i^{VU}$) is introduced to represent power quality externalities that are not captured by energy, loss, or congestion costs alone. When computing the DLMPs, the marginal cost associated with VU penalties ($c_i^{VU}$) serves as a coefficient reflecting these external costs. This penalty coefficient is treated as a tunable parameter that encodes the assumed economic valuation of power quality degradation (e.g., asset aging–related costs). Although several studies have examined asset lifetime impacts and the economic consequences of voltage unbalance~\cite{Jalilian-Induction, Molzahn_rep}, they do not result in a clear or universally accepted pricing scheme. More importantly, establishing such a pricing framework lies outside the scope of this study.

In the absence of standardized monetization rules for VU at the distribution level, values of $c_i^{VU} = 1$ and $1.5$ (relative to the marginal energy cost) are selected to reflect a comparable order of magnitude while avoiding unrealistic dominance in the welfare objective. As extensively discussed in previous sections, $c_i^{VU} = 1$ is not sufficient to reduce the VUF below the grid code threshold, whereas $c_i^{VU} = 1.5$ successfully achieves this.

\begin{figure}[!t]
    \centering
        \includegraphics[width=1.0\linewidth]{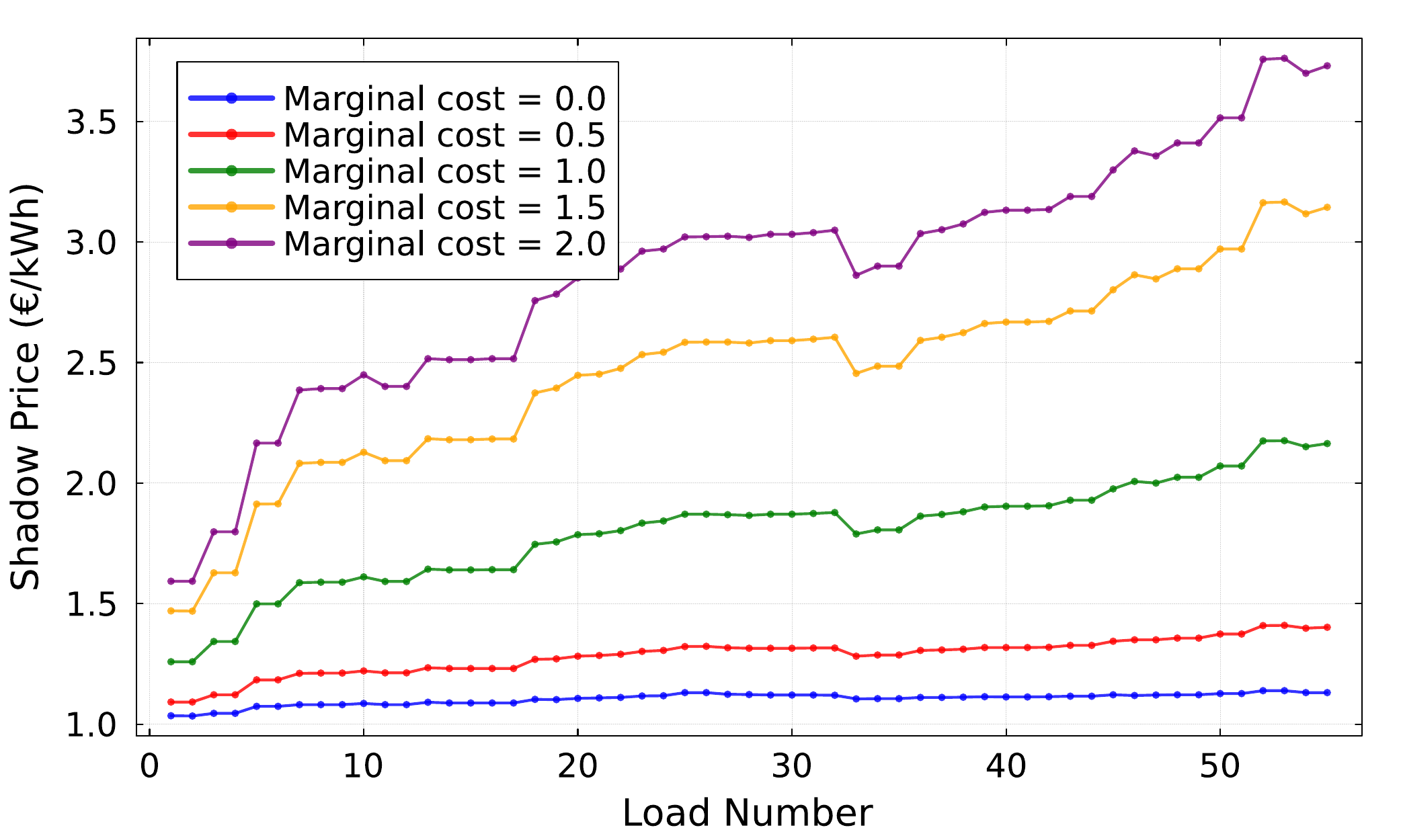}
    \caption{Sensitivity analysis results for phase `a'.}
    \label{fig:sens_anal}
\end{figure}

To further assess robustness, a sensitivity analysis is conducted over a practical range of $c_i^{VU}$ values from 0 to 2 with increments of 0.5. This analysis verifies that the qualitative conclusions remain consistent across reasonable penalty magnitudes. The corresponding results are summarized in Table~\ref{tab:sens_anal} and illustrated in Fig.~\ref{fig:sens_anal}.

\begin{table}[!t]
    \captionsetup{justification=centering, textfont={sc,footnotesize}, labelfont=footnotesize, labelsep=newline} 
    \renewcommand{\arraystretch}{1.2}
\centering
\caption{Sensitivity analysis over $c_i^{VU}$ results.}

\begin{tabular}{>{\centering\arraybackslash}p{1.5cm}
                >{\centering\arraybackslash}p{1.5cm}
                >{\centering\arraybackslash}p{1.8cm}
                >{\centering\arraybackslash}p{1.5cm}}
\toprule
{\footnotesize \textbf{Test case number}} & 
{\footnotesize \textbf{Total gen. cost (€)}} & 
{\footnotesize \textbf{Total losses (kWh})} & 
{\footnotesize \textbf{Highest VUF (\%)}} \\
\midrule
0.0 & 434.7 & 6.15 & 1.13 \\
0.5 & 434.7 & 6.15 & 1.13 \\
1.0 & 434.9 & 5.67 & 1.10 \\
1.5 & 438.1 & 4.49 & 1.00 \\
2.0 & 450.2 & 5.17 & 0.71 \\
\bottomrule
\end{tabular}

\label{tab:sens_anal}
\end{table}

\section{Conclusion}
\label{sec:conclusion}
This paper introduces a novel formulation and decomposition of distribution locational marginal prices that explicitly accounts for voltage unbalance. By embedding voltage unbalance penalties within a three-phase social-welfare optimization framework, the value of penalty is adjustable, the proposed approach generates economic signals that quantify the cost of unbalance and reveal its influence on operational decisions.

The study compares two distinct philosophies: \textit{hard limits}, which enforce strict compliance with technical standards, and \textit{soft limits}, which internalize the cost of unbalance through penalty terms in the objective function. This comparison highlights how these approaches lead to different DLMP patterns and cost signals, potentially guiding operational practices and investment strategies. While the \textit{soft limit} method demonstrates superior numerical behavior and scalability, the \textit{hard limit} approach remains essential for scenarios requiring strict adherence to grid codes. 

Case studies on two representative distribution networks under multiple scenarios reveal important insights. For example, the phase with the highest load imbalance does not always incur the largest excessive cost; in several cases, the middle phase receives the highest excessive cost. Furthermore, under certain conditions, adding load can reduce unbalance and lower overall system costs, creating opportunities for load-side participation and even revenue generation.

Overall, the unbalance-aware DLMP framework provides a theoretical foundation for analyzing the economic consequences of voltage unbalance. Future work may focus on improving scalability or exploring how penalty design interacts with asset degradation models and customer preferences to achieve socially optimal solutions.

\section*{Acknowledgments}

This work was supported by MICIU/AEI/10.13039/501100011033 and ERDF/EU under grant PID2022-141609OB-I00, and by the Madrid Government (Comunidad de Madrid-Spain) under the Multiannual Agreement 2023-2026 with Universidad Politécnica de Madrid, `Line A - Emerging PIs' (grant 24-DWGG5L-33-SMHGZ1).
The work of Alireza Zabihi was supported by the 2023 FPI-UPM call for Predoctoral Contracts within the framework of the 2021-2023 State Plan for Scientific, Technical, and Innovative Research.

\bibliographystyle{elsarticle-num}
\balance
\bibliography{Bibliography}

\end{document}